\documentclass{ws-procs9x6}

\begin{document}

\title{Domains in the Nonperturbative QCD Vacuum}

\author{ALEXANDER C. KALLONIATIS\footnote{\uppercase{S}upported 
by the \uppercase{A}ustralian \uppercase{R}esearch \uppercase{C}ouncil.}}

\address{CSSM, University of Adelaide, \\ 
Adelaide, 5005, South Australia, Australia\\ 
E-mail: akalloni@physics.adelaide.edu.au}

\author{SERGEI N. NEDELKO\footnote{\uppercase{P}artially 
supported by \uppercase{RFBR}
grant 01-02-17200.}}

\address{Bogoliubov L. of Theoretical Physics, JINR,\\ 
141980 Dubna, Russian Federation\\
E-mail: nedelko@thsun1.jinr.ru}

\maketitle

\abstracts{
QCD vacuum gluon fields are modelled by  
hyperspherical domains of constant field strength.
The pure glue theory confines static quarks. 
Solutions of the Dirac operator have chirality properties 
in agreement with lattice results.}

Long-range gluon fields in the QCD vacuum 
are suspected to be sources for 
confinement and chiral-symmetry breaking.
We proposed\cite{NKYM01} a Euclidean-space model for such fields
as idealised domains of hyperspherical geometry 
enclosing regions of covariantly constant Abelian 
self-dual or anti-self-dual field strength\cite{leut}.
The domains are characterised by two parameters, $R$ and $B$, the
average domain size and field strength magnitudes respectively.
We assign the following boundary conditions to 
gluon $A_{\mu}$ and quark $\psi$ fluctuations
respectively at the $j$-th domain:   
$\breve n_{(j)} A^{(j)}_\mu  =  0$,
and  
\begin{equation}
\label{bc}
\psi=i\!\not\!\eta^{(j)} e^{i\alpha_j\gamma_5}\psi,
\bar\psi=-\bar\psi i\!\not\!\eta^{(j)} e^{-i\alpha_j\gamma_5},
\end{equation} 
for quarks, with antihermitian $\gamma$ matrices and 
$\gamma_5=\gamma_1 \gamma_2 \gamma_3 \gamma_4$. These conditions can arise
from the presence of finite action singular fields at the boundaries. 
The adjoint matrix $\breve n_{(j)}=T^an_{(j)}^a$, $n_{(j)}^a$
a constant unit vector selecting the Cartan subalgebra, 
appears in the condition 
for gluons, and for the quarks   
the unit radial vector $\eta^{(j)}_\mu(x)=x_{\mu}/|x|$ is normal to the $j$-th
surface. The covariantly constant (anti-)-self-dual mean field 
\pagestyle{empty}
$ F^{a}_{\mu\nu}(x)
=\sum_{j=1}^N n^{(j)a}B^{(j)}_{\mu\nu}\theta(1-(x-z_j)^2/R^2)$,
with $B^{(j)}_{\mu\nu}B^{(j)}_{\mu\rho}=B^2\delta_{\nu\rho}$
in each domain, compensates for a decoupled treatment of domains.

To zeroeth order in gluon fluctuations the model gives
a gluon condensate density   
$g^2 \langle F^a_{\mu\nu}(x)F^a_{\mu\nu}(x)\rangle=4B^2,$
topological charge per domain $q={B^2R^4}/{16},$
topological susceptibility
$\chi = {{B^4 R^4}/{128 \pi^2}}$, 
and an $SU(3)_{{\rm colour}}$ string tension of the form  
$\sigma=Bf(BR^2)$ with the function  
\begin{eqnarray}
\label{sig-su3}
f(z)=\frac{2}{3\pi z}
\left(3-
\frac{\sqrt{3}}{2\pi z}\int_0^{2\pi z/\sqrt{3}}\frac{dx}{x}\sin x
 - \frac{2\sqrt{3}}{\pi z}\int_0^{\pi z/\sqrt{3}}\frac{dx}{x}\sin x
\right)
\nonumber
\end{eqnarray}
which is positive for $z>0$ and has a maximum for $z= 1.55$.
Estimating the model parameters at this $z-$value 
we fit the string constant to the lattice result,
and obtain $\sqrt{B}=947{\rm MeV}, \ R^{-1}=760 {\rm MeV}$,
and thus get for the gluonic parameters of the vacuum
$\sqrt{\sigma}=420 {\rm MeV}, \ \chi=(197 {\rm MeV})^4,
\ \frac{\alpha_s}{\pi}\langle F^2\rangle=0.081({\rm GeV})^4$
and $q=0.15$, while the density of the system is $42 \ {\rm fm}^{-4}$.
An area law does not occur for adjoint charges 
because of the presence of 
zero eigenvalues of the adjoint matrix $n^aT^a$.

The massless Dirac operator in a domain,
\begin{eqnarray}
(i\!\not\!D-\lambda)\psi(x)=0,
D_\mu=\partial_\mu + \frac{i}{2}\hat n B_{\mu\nu}x_\nu,
\nonumber
\end{eqnarray}
subject to the boundary condition Eq.~(\ref{bc}), is 
exactly soluble\cite{NKferm02}. The matrix $\hat n$ is in the fundamental
representation.  
We decompose eigenspinors via  
\begin{eqnarray}
\psi_{\kappa}=i\!\!\not\!\eta \chi_{\kappa} +\varphi_{\kappa},
\gamma_5\chi_{\kappa}=\mp\chi_{\kappa},
\gamma_5\varphi_{\kappa}=\mp\varphi_{\kappa},
\nonumber
\end{eqnarray}
where 
$\chi_{\kappa} = -\frac{1}{i\lambda_{\kappa}} 
\not\!\eta \not\!\!D\varphi_{\kappa}$, and 
$\not\!\!D^2\varphi_{\kappa}  =  \lambda_{\kappa}^2 \varphi_{\kappa}$. 
To give the quantum numbers $\kappa$ we use projectors 
$ N_\pm=\frac{1}{2}(1\pm \hat n/|\hat n|)$
and 
$\Sigma_\pm=\frac{1}{2}(1\pm \vec\Sigma\vec B/B)$
with $\hat B=|\hat n|B$ and the mixed projector
$O_{\zeta}=N_+\Sigma_{\zeta} + N_-\Sigma_{-\zeta}$ with $\zeta=\pm$.
Then $\kappa=(\zeta,n,k,m_{1,2})$.
Eq.(\ref{bc}) gives $\chi_{\kappa}=e^{\mp i \alpha} \varphi_{\kappa}$
on the boundary if the domain is (anti-)self-dual, possible  
only if the chiralities of the spinors $\chi$ and $\varphi$ 
are consistent with the duality of the vacuum gluon field: 
$\gamma_5 \varphi=\mp \varphi$ and $\gamma_5 \chi=\mp \chi$ if the field
$B_{\mu \nu}$ is (anti-)self-dual.
The discrete eigenvalues $\lambda_{\kappa}$ include then a principal quantum
number $n=1,2,\dots$. The 
angular momentum orbital and azimuthal quantum numbers
take values $k=0,1,\dots$ and $m_{1,2}=-k/2, \dots, k/2$.  
Zero modes $\lambda=0$ are absent.

\begin{figure}[htb]
\includegraphics*{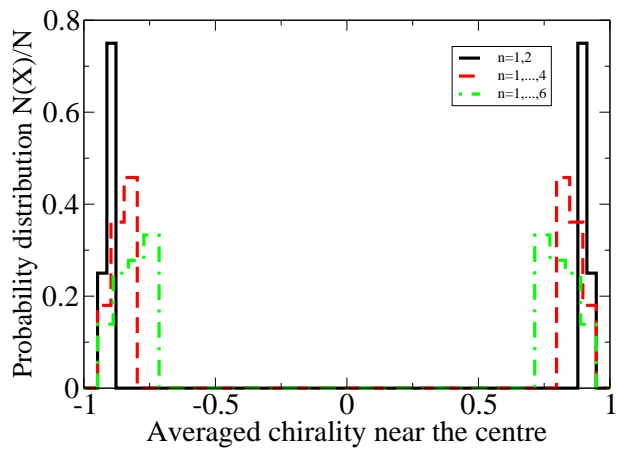}
\caption{
Histogram of chirality parameter $\bar X$
averaged over the central region with radius $0.025$fm.
Plots given in solid, dashed and dot-dashed lines incorporate all modes with
$n\le2$, $n\le4$ and  $n\le6$ respectively.}
\label{fig:X-hist}
\vspace*{-8mm}
\end{figure}

As an application, we study the local chirality parameter 
$X(x)$ defined\cite{Hor} via  
$\tan \left( \frac{\pi}{4}(1 - X(x)) \right) = 
\sqrt{ {\psi^{\dagger}(x)(1-\gamma_5)\psi(x)} \over 
{\psi^{\dagger}(x)(1+\gamma_5)\psi(x)} }$
which gives $X(x)=\pm 1$ at $x$  
where $\gamma_5 \psi(x) = \pm \psi(x)$. 
For lattice overlap fermions, histograms of $X$ measured at  
maxima of $\psi^{\dagger} \psi$
for low-lying overlap-Dirac eigenmodes 
peak\cite{EH01} at $X\approx \pm 1$ suggesting that  
low-lying  modes are strongly chiral and    
filter the duality of ``lumps'' underlying the
gauge field fluctuations. 
At domain centres radial modes are exactly chiral and probability densities
are maximal. The ``width'' of the peaks at half-maximum for these  
modes is $\alpha$-dependent and of the order $0.12-0.14 {\rm{fm}}$. 
The chirality of the modes are characterised via a histogram  
by averaging $X(x)$ over a small neighborhood of the domain centre.
With all $\alpha$
equally probable, we compute the probability to find a given value
of smeared $X$ among a set of modes. The result in Fig.~\ref{fig:X-hist}
is for the lowest modes and all spin-color orientations. 
The double peaks at $X\approx \pm 0.8$ 
and the above values for width and density of domains are  
consistent with the lattice\cite{EH01,Hor2}.  
This is argued to signal spontaneously broken chiral symmetry
on the lattice.
In our model this is achieved
{\it without} fermionic zero modes; strongly chiral low-lying
modes suffice.
Explicit calculation of the chiral condensate in the 
model is underway, though an estimate of the condensate density at
the domain centre\cite{NKYM01} gives $-(228 {\rm {MeV}})^3$.

\end{document}